\documentclass[12pt, A4]{article}
\usepackage{amssymb}

\begin{document}
\date{}

\title{Proxy Signature Scheme with Effective Revocation Using Bilinear Pairings}
\author{Manik Lal Das$^{1, 2}$\thanks{Corresponding author.}, Ashutosh Saxena$^1$, Deepak B.
Phatak$^2$\\\\
$^1$Institute for Development and Research in Banking Technology\\
Castle Hills, Masab Tank, Hyderabad-500057, India.\\
\texttt{Email:\{mldas, asaxena\}@idrbt.ac.in}\\\\
$^2$K. R. School of Information Technology\\
Indian Institute of Technology, Bombay\\
Mumbai-400076, India.\\
\texttt{Email:\{mdas, dbp\}@it.iitb.ac.in}}

\maketitle

\begin{abstract}
We present a proxy signature scheme using bilinear pairings that
provides effective proxy revocation. The scheme uses a
binding-blinding technique to avoid secure channel requirements in
the key issuance stage. With this technique, the signer receives a
partial private key from a trusted authority and unblinds it to
get his private key, in turn, overcomes the key escrow problem
which is a constraint in most of the pairing-based proxy signature
schemes. The scheme fulfills the necessary security requirements
of proxy signature and resists other
possible threats.\vspace{2 mm}
\end{abstract}

\section{Introduction}
Proxy signature is a digital signature where an original signer
delegates his signing capability to a proxy signer, and then the
proxy signer performs message signing on behalf of the original
signer. The notion of proxy signature has been evolved over a long
time, 16 years now \cite{gas89}. However, the cryptographic
treatment on proxy signature was introduced by Mambo et al
\cite{mam96} in 1996. They classified the delegation capability in
three types, namely \emph{full delegation}, \emph{partial
delegation} and \emph{delegation by warrant}. In full delegation,
an original signer directly gives his private key to a proxy
signer and using it the proxy signer signs the document. The
drawback of proxy signature with full delegation is that the
absence of a distinguishability between the original signer and
the proxy signer. In partial delegation, the original signer
derives a proxy key from his private key and hands it over to the
proxy signer as a delegation of signing rights. In this case, the
proxy signer can misuse the delegation of signing rights, because
partial delegation does not restrict the proxy signer's signing
capability. The weakness of full and partial delegations are
eliminated by partial delegation with warrant, where a warrant
explicitly states the signers' identity, delegation period and the
qualification of the message on which the proxy signer can sign,
etc. Once proxy delegation is given, the revocation is an
important issue in the proxy signature scheme. For instance, the
original signer key is compromised or any misuse of delegation of
signing rights is noticed. It may so happen that the original
signer wants to terminate his delegation power before the expiry
e.g., the manager of a company has come
back from his trip before time that he was scheduled for.\\
Desirable security properties of proxy signatures have evolved
over this period and a widely accepted list of required properties
are as follows:
\newcounter{a}
\begin{list}{-}
{\usecounter{a}} \item Strong unforgeability: A designated proxy
signer can create a valid proxy signature on behalf of the
original signer. But the original signer and other third parties
cannot create a valid proxy signature. \item Strong
identifiability: Anyone can determine the identity of the
corresponding proxy signer from the proxy signature. \item
Verifiability: The verifier can be convinced of the original
signer's agreement from the proxy signature. \item
Distinguishability: Proxy signatures are distinguishable from
normal signatures by everyone. \item Strong undeniability: Once a
proxy signer creates a valid proxy signature, he cannot deny the
signature creation. \item Prevention of misuse: The proxy signer
cannot use the proxy key for other purposes than it is made for.
That is, he cannot sign message with the proxy key that have not
been defined in the warrant. If he does so, he will be identified
explicitly from the warrant.
\end{list}
After Mambo et al.'s \cite{mam96} scheme, several schemes have
been proposed \cite{kim97}, \cite{oka99}, \cite{sun00},
\cite{bol03}, \cite{her04}. However, most of the schemes lack
proxy revocation mechanism. Recently, the bilinear pairings,
namely the Weil pairing and the Tate pairing of algebraic curves
have been found important applications \cite{bon01}, \cite{coc01},
\cite{hes02} in identity(ID) based cryptography. The advantage of
an ID-based cryptography \cite{sha84} is that it avoids public key
certification, the public key of a user is his identity, e.g.,
e-mail, social security number, etc. There are a few proxy
signature schemes \cite{che03}, \cite{zha03}, \cite{xu04},
\cite{zha04} based on bilinear pairings; however, the schemes lack
the key escrow problem and have not addressed the proxy revocation
mechanism. In this paper, we present a proxy signature scheme
using bilinear pairings that provides effective proxy revocation
mechanism. Our scheme is not exactly ID-based, it is a variant of
ID-based schemes. The scheme does not require secure channel in
the key issuance stage and avoids the key escrow problem.\\ The
rest of the paper is organized as follows. Section 2 discusses
some preliminaries. Section 3 presents the scheme. Section 4
analyzes the security and performance of the scheme. Finally, we
conclude the paper in Section 5.

\section{Preliminaries}
\subsection{Bilinear Pairings}
Suppose $G_1$ is a cyclic additive group of prime order $q$,
generated by $P$, and $G_2$ is a cyclic multiplicative group of
the same order $q$. A map $\hat{e} : G_1 \times G_1 \rightarrow
G_2$ is called a bilinear mapping if it satisfies the following
properties:\\
\hspace*{2 mm}- Bilinear: $\hat{e}(aP, bQ) = \hat{e}(P,Q)^{ab}$
for all $P,Q \in G_1$ and $a, b \in \mathbb Z_q^*$ ;\\
\hspace*{2 mm}- Non-degenerate: There exist $P, Q \in G_1$
such that $\hat{e}(P,Q) \ne 1$ ;\\
\hspace*{2 mm}- Computable:  There is an efficient algorithm to
compute $\hat{e}(P,Q)$ $\forall$ $P, Q \in G_1$.\\
In general, $G_1$ is a group of points on an elliptic curve and
$G_2$ is a multiplicative subgroup of a finite field.

\subsection{Computational Problems}
Definition 1. Discrete Logarithm Problem (DLP) : Given $Q, R \in
G_1$, find an integer $a \in \mathbb Z_q^*$ such that
$R=aQ$.\vspace{2 mm}\\
Definition 2. Decisional Diffie-Hellman Problem (DDHP) : Given
$(P, aP, bP, cP)$ for $a, b, c \in \mathbb Z_q^*$, determine
whether $c \equiv ab$ mod $q$. The advantage \texttt{Adv} of any
probabilistic polynomial-time algorithm $\mathcal{A}$ in solving
DDHP in $G_1$ is defined as: \[\texttt{Adv}^{DDH}_{{\mathcal{A}},
G_1} = \left[\texttt{Pr}[{\mathcal{A}}(P, aP, bP,
cP)=1]-\texttt{Pr}[{\mathcal{A}}(P, aP, bP, abP)=1]: a,b,c \in
\mathbb Z_q^*\right].\] For every probabilistic polynomial-time
algorithm $\mathcal{A}$, \texttt{Adv}$_{{\mathcal{A}}, G_1}^{DDH}$
is negligible.\vspace{2 mm}\\
Definition 3. Computational Diffie-Hellman Problem (CDHP) :
Given $(P, aP, bP)$ for $a, b \in \mathbb Z_q^*$, compute $abP$.\\
The advantage of any probabilistic polynomial-time algorithm
$\mathcal{A}$ in solving CDHP in $G_1$ is defined as:
\[\texttt{Adv}_{{\mathcal{A}}, G_1}^{CDH} =
\left[\texttt{Pr}[{\mathcal{A}}(P, aP, bP, abP)=1: a,b \in \mathbb
Z_q^*\right].\] For every probabilistic algorithm
$\mathcal{A}$, \texttt{Adv}$_{{\mathcal{A}}, G_1}^{CDH}$ is negligible.\vspace{2 mm}\\
Definition 4. Gap Diffie-Hellman Problem (GDHP): A class
of problems where DDHP is easy while CDHP is hard.\vspace{2 mm}\\
Definition 5. Weak Diffie-Hellman Problem (WDHP) : Given $(P, Q,
aP)$ for $a \in Z_q^*$, compute $aQ$.

\section{The Proposed Scheme}
To avoid the original signer's forgery  and prevention of
delegation power misuse, the proxy-protected proxy signature
\cite{kim97} is a secure approach. Our scheme is based on
proxy-protected notion and uses the merits of partial delegation
with warrant\footnote{A warrant consists of original signer and
proxy signer identities, qualification of the message on which the
proxy signer can sign, validity period of the
delegation, etc.}.\\
The participating entities and their roles in the proposed scheme
are defined as follows:
\newcounter{c}
\begin{list}{$\bullet$}
{\usecounter{c}} \item Private Key Generator (PKG): A trusted
authority who receives signer's identity (ID) along with other
parameters, checks validity of ID and issues partial private key
to the signer corresponding to the ID. \item Original Signer:
Entity who delegates his signing rights to a proxy signer. \item
Proxy Signer: Entity who signs the message on behalf of the
original signer. \item Verifier: Entity who verifies the proxy
signature and decides to accept or reject.
\end{list}
The scheme has five phases: Setup, KeyGen, ProxyKeyGen,
ProxySignGen and ProxySignVerify.
The phases work as follows.\\
\texttt{[ Setup ]}\\
It takes as input a security parameter; and outputs system
parameters \textit{params} and master-key of PKG. The
\textit{params} includes a cyclic additive group $G_1$ of prime
order $q$ generated by $P$, a cyclic multiplicative group $G_2$ of
prime order $q$, a bilinear map $\hat{e} : G_1 \times G_1
\rightarrow G_2$, hash functions $H_1 : \{0,1\}^* \times G_1
\times G_1 \rightarrow G_1$, $H_2 : \{0,1\}^* \rightarrow G_1$, $h
: \{0,1\}^* \times G_1 \times G_1 \rightarrow \mathbb Z_q^*$, and
public key of PKG. The PKG selects a master-key $s \in \mathbb
Z_q^*$ and computes public key as $Pub_{PKG} = sP$. The PKG
publishes \textit{params} = $(G_1, G_2, \hat{e}, q, P, Pub_{PKG}, H_1, H_2, h)$ and keeps $s$ secret.\vspace{1 mm}\\
\texttt{[ KeyGen ]}\\
It takes user chosen parameters and \textit{params} as inputs; and
outputs user private key. The entire phase consists of a
\textit{partial private key issuance} and a \textit{private key
generation} stages. The stages use a binding-blinding technique to
avoid the key escrow problem and to eliminate the secure channel
requirements. The binding-blinding technique works as follows:
\newcounter{dd}
\begin{list}{-}
{\usecounter{dd}} \item The user chooses two secret binding
factors, calculates the binding parameters and sends them to the
PKG over a public channel along with his/her identity. \item As
the communication channel between the user and the PKG is a public
channel, a dishonest party can construct his/her preferred binding
parameters using the targeted user's identity and sends the
binding parameters along with user's identity before the user
submits a request for partial private key. To avoid this type of
attack, the PKG first sends a message to the email-id\footnote{At
this juncture, we assume that the email-id acts as the user
identity; however, other identity could play the same role if it
avoids the unregistered identity attack. We note that it is a
difficult task to avoid the unregistered identity attack for any
types of identity if there is no off-line (secure channel)
interaction between the PKG and the user, in turn it opens a
prominent future scope of our proposed work.} (email-id acts as
the user identity) and asks a confirmation from the email-id
owner. If the email-id owner confirms his/her request for a
partial private key, then the PKG proceeds to the next step. \item
The PKG checks the validity of binding parameters. Upon successful
validation of the parameters, the PKG computes signer partial
private key. Then, the PKG sends the partial private key to the
user in a blinding manner over the public channel.
\end{list}
\texttt{PartialPrivateKey} issuance:
\newcounter{d}
\begin{list}{-}
{\usecounter{d}} \item User $U_{\lambda}$ computes his own public
key $Pub_{\lambda} = H_2(ID_{\lambda})$. \item $U_{\lambda}$ picks
two secret binding factors $a_{\lambda}, b_{\lambda} \in \mathbb
Z_q^*$ and computes $X_{\lambda}=a_{\lambda}Pub_{\lambda}$,
$Y_{\lambda}=a_{\lambda}b_{\lambda}Pub_{\lambda}$,
$Z_{\lambda}=b_{\lambda}P$ and
$W_{\lambda}=a_{\lambda}b_{\lambda}P$. Then he sends
$(X_{\lambda}$, $Y_{\lambda}$, $Z_{\lambda}$, $W_{\lambda}$,
$ID_{\lambda})$ to the PKG over a public channel. \item Once the
$ID_{\lambda}$ is correct (we assume that identity of the user is
his/her email-id and unregistered identity attack can be avoided
by the above mentioned email confirmation procedure), the PKG
computes $Pub_{\lambda} = H_2(ID_{\lambda})$ and verifies the
validity of $ID_{\lambda}$ by whether $\hat{e}(Y_{\lambda}, P) =
\hat{e}(X_{\lambda}, Z_{\lambda}) = \hat{e}(Pub_{\lambda},
W_{\lambda})$. \item The PKG computes $U_{\lambda}$'s partial
private key as $D_{\lambda} = sY_{\lambda}$ and creates a
\textit{registration-token} $Reg_{\lambda} = sZ_{\lambda}$
corresponding to $ID_{\lambda}$. Then, PKG publishes
$(Reg_{\lambda}, ID_{\lambda})$ in a public directory and sends
$D_{\lambda}$ to $U_{\lambda}$ over a public channel.
\end{list}
We note that the PKG controls the public directory and checks
every request before issuance of any partial private key. If the
identity is present in the directory, the PKG denies the request,
thereby the registration-token replacement is not possible by any
other party.\vspace{1 mm}\\
\texttt{PrivateKey} generation:
\newcounter{d1}
\begin{list}{-}
{\usecounter{d1}} \item On receiving the partial private key
$D_{\lambda}$, the signer $U_{\lambda}$ checks its correctness by
whether $\hat{e}(D_{\lambda}, P) = \hat{e}(Y_{\lambda},
Pub_{PKG})$. If $D_{\lambda}$ is valid, $U_{\lambda}$ unblinds it
and generates his private key as $S_{\lambda} =
a_{\lambda}^{-1}D_{\lambda}$.
\end{list}
\underline{Original signer private key}: Let $ID_o$ be the
identity of an original signer. The original signer chooses
binding secret factors $a_o$ and $b_o$ and runs the
\texttt{KeyGen} algorithm to get his partial private key
as\vspace{1 mm}\\\hspace*{10 mm} $D_o$ $\leftarrow$
\texttt{PartialPrivateKey}$(X_o, Y_o, Z_o, W_o, ID_o)$.\vspace{1 mm}\\
After validating $D_o$, the original signer generates his private key as $S_o=a_o^{-1}D_o$.\vspace{1 mm}\\
\underline{Proxy signer private key}: Let $ID_p$ be the identity
of the proxy signer. The proxy signer chooses the binding factors
$a_p$ and $b_p$ and runs the \texttt{KeyGen} algorithm to get his
partial private key as\vspace{1 mm}\\\hspace*{10 mm} $D_p$
$\leftarrow$ \texttt{PartialPrivateKey}$(X_p, Y_p, Z_p, W_p,
ID_p)$.\vspace{1 mm}\\
After validating $D_p$, the proxy signer generates his private key as $S_p=a_p^{-1}D_p$.\vspace{1 mm}\\
\texttt{[ ProxyKeyGen]}
\newcounter{dd1}
\begin{list}{-}
{\usecounter{dd1}} \item The original signer and proxy signer
agree on a warrant $m_w$. \item The original signer computes $U_o
= S_o + b_oH_1(m_w, Pub_o, Pub_p)$, $\psi_o = b_oP$ and sends the
tuple $(m_w, U_o, \psi_o, Pub_o)$ to the proxy signer over a
public channel as the delegation capability. \item The proxy
signer checks whether\\\hspace*{10 mm} $\hat{e}(U_o, P) =
\hat{e}(\psi_o, H_1(m_w, Pub_o, Pub_p))\hat{e}(Pub_o, Reg_o)$.
\item If the delegation capability is valid, the proxy signer
computes proxy key as\\\hspace*{10 mm} $V_p = U_o + S_p +
b_pH_1(m_w, Pub_o, Pub_p)$.
\end{list}
\vspace{1 mm}
\texttt{[ ProxySignGen ]}\\
To sign a message $m$, the proxy signer computes the following
steps:
\newcounter{e}
\begin{list}{-}
{\usecounter{e}} \item Select a random $r \in \mathbb Z_q^*$ and
computes $R = rP$. \item Compute $a = h(m, R, Pub_p)$ and $\psi_p
= b_pP$. \item Compute $V = (r+a)^{-1}V_p$.\end{list} The proxy
signature on $m$ is the tuple $(m_w, m, R, V, \psi_o, \psi_p,
Pub_o, Pub_p)$.\vspace{1 mm}\\
\texttt{[ ProxySignVerify ]}\\
The proxy signature $(m_w, m, R, V, \psi_o, \psi_p, Pub_o, Pub_p)$
is valid if and only if\vspace{3 mm}\\\hspace*{3 mm} $\hat{e}(R +
h(m, R, Pub_p)P, V)$\\\hspace*{10 mm} $= \hat{e}(\psi_o + \psi_p,
H_1(m_w, Pub_o, Pub_p))\hat{e}(Pub_o, Reg_o)\hat{e}(Pub_p,
Reg_p)$.
\section{Analysis of the Scheme}
\subsection{Correctness of proxy signature verification}
$\hat{e}(R + h(m, R, Pub_p)P, V)$

= $\hat{e}((r + h(m, R, Pub_p))P, (r + a)^{-1}V_p)$

= $\hat{e}((r + a)P, (r + a)^{-1}V_p)$

= $\hat{e}(P, U_o + S_p + b_pH_1(m_w, Pub_o, Pub_p))$

= $\hat{e}(P, S_p + S_o + (b_p + b_o)H_1(m_w, Pub_o, Pub_p))$

= $\hat{e}(P, S_p)\hat{e}(P, S_o)\hat{e}(P, (b_p + b_o)H_1(m_w,
Pub_o, Pub_p))$

= $\hat{e}(Pub_o, Reg_o)\hat{e}(Pub_p, Reg_p)\hat{e}((b_o + b_p)P,
H_1(m_w, Pub_o, Pub_p))$

= $\hat{e}(Pub_o, Reg_o)\hat{e}(Pub_p, Reg_p)\hat{e}(\psi_o +
\psi_p, H_1(m_w, Pub_o, Pub_p))$

\subsection{Security Analysis}
In this section, we show that the proposed scheme satisfies the
security properties of a proxy signature, mentioned in Section 1.
In addition, the scheme withstands some other possible threats.\vspace{0.5 mm}\\
\emph{The scheme can withstand the strong unforgeability security
property.}\\
To create a valid proxy signature, one should need the original
signer and proxy signer private keys. Though the adversary can
intercept signer partial private key $D_i$ ( i.e.,
$sa_ib_iPub_i$), he cannot construct the private key $S_i$ (i.e.,
$sb_iPub_i$) without the knowledge of $a_i$, because it is a WDHP
(definition 5) which is assumed to be a hard problem. As our
scheme is proxy protected, i.e., the proxy signer has to use his
private key and original signer's delegation power to sign a
message, thus, the original signer is also prohibited from forging
a valid proxy signature. Moreover, the PKG cannot frame the
signers' with the knowledge of binding parameters ($X_i, Y_i, Z_i,
W_i$), as extracting the binding factors $a_i$, $b_i$ from the
binding parameters is as hard as CDHP (definition 3).\vspace{0.5 mm}\\
\emph{The scheme can resist the identifiability, undeniability and
distinguishability security properties.}\\
A valid proxy signature of a message $m$ is the tuple $(m_w$, $m,
R$, $V$, $\psi_o$, $\psi_p$, $Pub_o$, $Pub_p)$. The public keys
$Pub_o$, $Pub_p$ and warrant $m_w$ are the straightforward
witnesses (i.e., identities) of the signers. In addition, a
verifier will come to know the agreement between original and
proxy signers from $m_w$.\\
From the correctness of the proxy signature, given in Section 4.1,
it is clear that the proxy signer cannot deny his signature
creation. The verification of a valid proxy signature needs the
proxy signer's public key, in turn, proves that the signature was
created by the proxy signer. Further, the PKG can also prove the
identity of the proxy signer, as the tuple ($Reg_p, ID_p$) in the
PKG public directory is a supporting identification of a proxy
signer and is also required in the proxy signature verification
phase. Any verifier will receive the proxy signature that contains
warrant $m_w$ and the public key of signers, by which the verifier
can easily distinguish the proxy signature from the normal
signature.\vspace{0.5 mm}\\
\emph{The scheme is secure against misuse of the proxy delegation.}\\
In the Proxy key generation phase, the original signer signs the
tuple ($m_w$, $Pub_o$, $Pub_p$) and gives it to the proxy signer
as his delegation capability. The proxy signer signs a message
with the proxy key that is being created by his private key and
original signer's delegation capability. The qualification of
message and limitation of proxy is clearly defined in $m_w$ and
the delegation is made for the designated proxy signer only. If
the proxy signer misused the delegation capability, the proxy
signer will be detected by any verifier from $m_w$. The original
signer's misuse is also prevented because he cannot create a valid
proxy signature against the name of the proxy signer.\vspace{1 mm}\\
Apart from the above security properties, the scheme withstands
the following possible threats.\vspace{0.5 mm}\\
\emph{Threat 1. Registration-token replacement} : The PKG creates
\textit{registration-token} corresponding to each registered
signer and publishes it along with signer-ID in a public
directory, which is controlled by PKG only. If a request comes
from signer identity $ID^*$ for issuance of a partial private key,
the PKG first checks whether $ID^*$ is in the public directory. If
it is found in the public directory, the PKG rejects the request,
otherwise executes the \texttt{KeyGen} algorithm for $ID^*$. Thus,
the registration-token replacement is not possible by any party
(the PKG itself can replace the registration-token, but we
assume that the signer trusts PKG for not to do it).\vspace{0.5 mm}\\
\emph{Threat 2. Man-in-the-middle attacks} : In our scheme, the
communication channel of the key issuance stage is a public
channel, thus an attacker may try to calculate the private key or
binding factors of a signer by intercepting the binding parameters
and partial private key. On intercepting the binding parameters,
the adversary can formulate the following problem : \emph{Given
\textit{params}, binding parameters ($a_iPub_i$, $a_ib_iPub_i$,
$b_iP$, $a_ib_iP$, $ID_i$) and partial private key $D_i$ (i.e.,
$sa_ib_iPub_i$); Compute private key $S_i$ (i.e. $sb_iPub_i$) or
binding factors ($a_i$, $b_i$)}. To solve this problem, one has to
solve either the CDHP or the WDHP, which is assumed to be
computationally hard.\vspace{0.5 mm}\\
\emph{Threat 3. ONE partial private key $\rightarrow$ MANY private
keys} : The scenario of generating more than one private key from
a partial private key is nor possible, because the private key
$S_i$ (i.e. $sb_iPub_i$) and the registration-token $Reg_i$ are
linked by the secret binding factor $b_i$. If a signer generates
another private $S_i^*$ from $S_i$ and signs a message by $S_i^*$,
then the verification of the signature fails because the change
from $S_i$ to $S_i^*$ is not reflected in $Reg_i$. Thereby, the
signer cannot perform this type of attempt without being
detected.\vspace{1 mm}\\
\textit{\textbf{Theorem 1.} The proxy signature scheme is said to
be secure against adaptive chosen-massage attacks under random
oracle model if no polynomially bounded adversary (in $k$) has
non-negligible advantage (in $k$).}\\
\textit{Proof}: The proof of the theorem is ascertained by the
following challenger-adversary game.\\
\textbf{Setup}: A challenger $\mathcal{C}$ takes a security
parameter $k$ and runs the \texttt{Setup} phase as mentioned in
Section 3. Then $\mathcal{C}$ returns the resulting system
parameters \textit{params} to $\mathcal{A}$ and keeps
master-key $s$ with itself.\\
\textbf{Queries}: The adversary $\mathcal{A}$ issues adaptively
the queries $q_1, q_2,\cdots, q_m$ in any order for the following:\\
\texttt{ProxyKeyGen} query on $Pub_j$, where $j=1,\cdots, m$:\\
$\mathcal{C}$ runs the \texttt{ProxyKeyGen} phase and generates
proxy key $V_j$ using
$S_j$ and $b_j$ corresponding to $Pub_j$, and sends it to $\mathcal{A}$.\\
\texttt{ProxySignGen} query on $(Pub_j, M^{\prime})$:\\
$\mathcal{C}$ runs the \texttt{ProxyKeyGen} phase and generates
the proxy key $V_j$. Then, $\mathcal{C}$ signs the message
$M^{\prime}$ and returns the proxy signature $(\omega, M^{\prime},
R^{\prime}, V(M^{\prime}), \psi_o, \psi_j, Pub_o,
Pub_j)$ to $\mathcal{A}$.\\
\textbf{Guess}: $\mathcal{A}$ outputs a proxy signature for
message $M^*$, where $M^*$ did not appear in the
\texttt{ProxySignGen} query.\\
\textbf{Result}: $\mathcal{A}$ wins if his produced proxy
signature on $M^*$ is valid. The advantage of $\mathcal{A}$ in
attacking the scheme is defined to be the probability that
$\mathcal{A}$ produces a valid proxy signature in the game. We say
that our scheme is secure against adaptive chosen-message attacks
under random oracle model if no polynomially bounded adversary has
non-negligible advantage in this game.

\subsection{Performance}
\textit{Proxy revocation} : The revocation of delegation
capability (i.e., proxy revocation) is an important concern in any
proxy signature scheme. It is observed that the schemes
\cite{che03}, \cite{zha03}, \cite{xu04}, \cite{zha04} have not
addressed the proxy revocation issues, which is a practical
requirement. In our scheme, proxy revocation can be easily done by
revoking the registration-token from the PKG's public directory.
If the original signer wants to revoke his delegation of signing
rights, he sends a revoke-request tuple $(M_r$, $m_w$, $Rev$,
$Pub_o$, $Pub_p$, $\psi_o)$ to the PKG and proxy signer, where
$Rev = S_o + b_oH_1(M_r, Pub_o, Pub_p)$ and $M_r$ states the
identity of the signer along with the reason for proxy revocation.
The PKG first checks the authenticity and validity of the
revoke-request and if the request is valid then PKG revokes the
tuple $(Reg_o, ID_o)$ and $(Reg_p, ID_p)$ from the public
directory. We note that the proxy signer will not object if the
PKG removes $(Reg_p, ID_p)$ without his consent (the original
consent is with PKG), because if the delegation capability is no
longer authorized, the delegated proxy signer is no longer
required. The PKG validates the revoke-request as follows:\\
\hspace*{15 mm}$\hat{e}(Rev, P) = \hat{e}(S_o + b_oH_1(M_r, Pub_o,
Pub_p), P)$\\
\hspace*{32 mm}= $\hat{e}(sb_oPub_o + b_oH_1(M_r, Pub_o, Pub_p),
P)$\\
\hspace*{32 mm}= $\hat{e}(Reg_o, Pub_o)\hat{e}(H_1(M_r,
Pub_o, Pub_p), \psi_o)$\vspace{2 mm}\\
\textit{Key escrow} : In our scheme, the PKG issues a
\texttt{PartialPrivateKey} to the signer and with this the signer
computes his private key. The PKG is not having knowledge of
signer private key. To construct a private key from the partial
private key, one has to know the secret binding factor or has to
solve DLP. As the binding factor is retained with the signer only,
other party can not obtain signer private key because solving DLP
is a hard problem. Thus, our scheme avoid the key escrow problem,
which occurs in the schemes \cite{che03}, \cite{zha03},
\cite{xu04}, \cite{zha04}\vspace{2 mm}\\
\textit{No need of secure channel} : To eliminate the secure
channel in the key issuance stage, we used a binding-blinding
technique where the signer requests for a partial private key from
the PKG. We considered a simplest procedure to verify the
genuineness of signer's identity while partial private key
issuance. After validating the signer request, the PKG issues a
partial private key in a blinded manner. Finally, the signer
unblinds the partial private key to get his private key. This
binding-blinding technique avoids the secure channel in the key
issuance stage.

\section{Conclusion}
We proposed a proxy signature scheme using bilinear pairings that
provides effective proxy revocation. The scheme uses a
binding-blinding technique to eliminate the secure channel
requirements in the key issuance stage. We considered a mechanism
to avoid the unregistered identity attacks when identity is user's
email-id, though the mechanism does not provide a generic solution
for other types of identities. We leave this problem as a future
scope of the proposed work. Our scheme is not exactly ID-based
scheme; however, it avoids the key escrow problem, which remains
constraint in most of the existing pairing-based proxy signature
schemes. We showed that the scheme satisfied the security
requirements of a proxy signature and also withstood other
possible threats.

\bibliographystyle{plain}

\end{document}